# Dynamics of capillary coalescence and breakup: quasi-two-dimensional nematic and isotropic droplets

P. V. Dolganov[1], A. S. Zverev[1], K. D. Baklanova[1,2], and V. K. Dolganov[1]

[1] *Institute of Solid State Physics, Russian Academy of Sciences, 142432, Chernogolovka, Moscow district, Russia*
[2] *National Research University Higher School of Economics, Moscow, 101000 Russia*

For the first time we observed formation of small satellite droplets from the bridge at droplet coalescence. Investigations were made using a Hele-Shaw cell in the two-phase region at nematic-isotropic phase transition. In previous works on coalescence it was considered that before start of coalescence there exists a bridge between the outer fluid connecting regions on the two sides of the droplets (outer bridge). After start of coalescence a bridge connecting the two droplets appears (droplet bridge) and the outer bridge is broken. For the first time we have shown that there are coalescence processes when after start of coalescence both the droplet bridge and the outer bridge can exist. This cardinally changes the coalescence process. During the first coalescence stage the size of the outer bridge decreases, the size of the droplet bridge increases. During the second stage the outer bridge becomes unstable with pinch-off, formation of pointed end domains, secondary instability, splitting of pointed end domains and formation of satellite droplets. Our work connects two areas of fluid dynamics: coalescence and breakup with formation of satellite droplets.

## I. INTRODUCTION

The study of droplet transformation driven by surface tension is a classical direction of investigations in fluid physics. Cardinal transformations occur at coalescence of three-dimensional (3D) [1-5] or 2D [6-12] droplets or inversely when satellite droplets are formed from a jet or a bigger stretched droplet [13-20]. Coalescence and breakup belong to the class of free surface problems. These phenomena are of importance for different areas of physics, chemistry, biology and various technological applications. Droplet coalescence and breakup of droplets and filaments are opposite phenomena both leading to a change of topology. Coalescence and breakup have been theoretically studied analytically, using scaling approach, numerical calculations and molecular dynamics [16,21-28]. Although coalescence and jet breakup are well known and daily observed, many questions remain especially in the case of complex internal structure of coalescing objects.

A perspective direction of investigation is the study of quasi-2D coalescence in Hele-Shaw cells [29-33]. Oswald and Poy [31] proposed a unique system for investigation, namely the two-phase region at the nematic-isotropic phase transition. Nematic droplets are formed on cooling the liquid into the two-phase region. Droplets with orientational order allow to expand the subjects of investigations aiming to study not only the transformation of droplet shape but also the transformation of interior droplet structure at coalescence. One more important peculiarity of two-phase media is the smallness of interfacial energy between nematic and isotropic phases (surface tension $\gamma$ is about $10^{-5}$ N/m [31]). As a result characteristic times $\tau$ are relatively large so that no high-speed techniques are required. Oswald and Poy [31] investigated the shape of isotropic droplets at the final coalescence stage when a nearly ellipsoidal droplet relaxed to the circular shape.

In the current paper we investigated quasi-two-dimensional coalescence of droplets at isotropic-nematic and nematic-isotropic phase transitions. In the first case nematic droplets coalesce in outer isotropic liquid. At the "reverse" transition isotropic droplets coalesce in surrounding nematic. The material was in confined geometry of a flat Hele-Shaw cell. We found that transformation of bridge region at coalescence cardinally differs from earlier considered. In previous works on quasi-2D coalescence the following model was assumed. Before coalescence there is a bridge between the outer fluid connecting regions on the two sides of the droplets (outer bridge, Fig. 1(a)). As droplets approach each other [Fig. 1(b)], the coalescence starts when the outer bridge is broken in some point and a bridge between coalescing droplets appears (droplet bridge, Fig. 1(c)). Then the width of the droplet bridge increases [Fig. 1(d)]. So, there is only one bridge at any moment of time (outer bridge or droplet bridge, Fig. 1). We found that there are coalescence processes where both the droplet bridge and the outer bridge can exist after start of coalescence. This cardinally changes the coalescence process with respect to known process. With time the size of the outer bridge decreases, the size of the droplet bridge increases. Our second finding is that when the width of the outer bridge becomes too small it becomes unstable. The instability leads to spontaneous pinch-off, formation of pointed end droplets, and second instability with generation of satellite droplets. To our knowledge, our findings are the first example of such behavior in droplet coalescence.

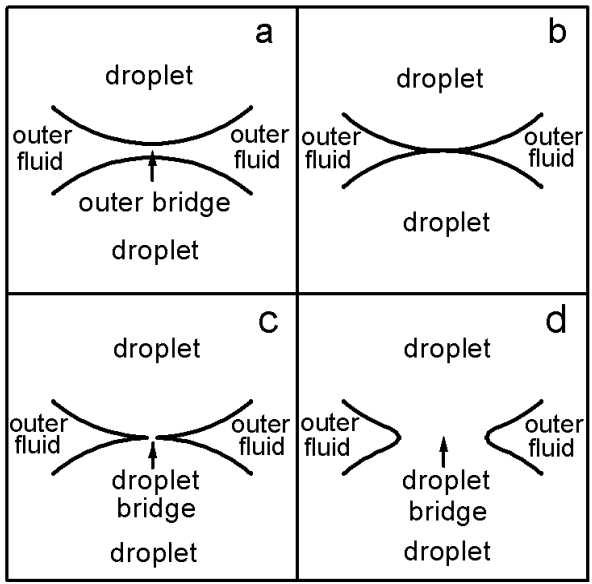

FIG. 1. Schematic illustration of conventional model of quasi-2D droplet coalescence in outer fluid. Droplets before coalescence with outer bridge (a,b) and during coalescence with the droplet bridge (c,d). The outer and droplet bridges are perpendicular to each other.

## II. EXPERIMENTAL DETAILS

We used the liquid crystal E7 (Synthon Chemicals) with the nematic-isotropic phase transition. Our samples have a two-phase region about $\Delta T \approx 2.2°C$ from about $T_{solidus} \approx 56.6°C$ to $T_{liquidus} \approx 58.8°C$. Wide two-phase region allows to prepare nematic or isotropic droplets and to observe their coalescence. The material was confined between two glass plates. The orientation of nematic director was parallel to the plates. We used cells with in-plane dimensions 1 cm × 1 cm. Sample thicknesses ranged from $h=5$ μm to $h=50$ μm. The experiments are performed using Olympus BX51 microscope. Optical observations were made in transmission between crossed polarizers. Droplet transformation was captured at a frame rate from 400 to 50 fps using Baumer VCXU-02C and Nikon D3300 video cameras.

Experiments were performed as follows: the sample was heated to a temperature somewhat higher than the two-phase region. For the investigations of coalescence of nematic droplets the sample was then cooled from high temperature. When nematic droplets appeared, cooling was stopped or was sufficiently slow (slower than 0.1°/min). Coalescence starts when droplets due to increase of their sizes come into contact. We selected droplets with approximately the same size (difference in radii $R$ was less than 8%). For investigation of coalescence of isotropic droplets the sample after cooling was heated to a temperature slightly lower than the two-phase region. Then the protocol was similar as for coalescence of nematic droplets but with heating which induced nucleation of isotropic droplets.

## III. RESULTS AND DISCUSSION

Figure 2 shows the dynamics of transformation of the nematic droplet shape and structures inside droplets during coalescence in isotropic environment. Two circular droplets [Fig. 2(a)] transform to a single waisted droplet [Fig. 2(b,c)], then to ellipsoidal [Fig. 2(d,e)] and circular [Fig. 2(f)]. Dynamics of the transformation will be described in subsection (a). Unusual and unexpected phenomena are formation of the line domain between coalescing droplets [the yellow stripe in Fig. 2(b)], its transformation to elongated [Fig. 2(c)] and circular domain [Fig. 2(d-f)]. With time or on cooling the size of the circular domain decreases and then it vanishes transforming to nematic. This can indicate the existence of the isotropic phase in the place of domains. Detailed description of the transformation of structure inside the droplets will be given in subsection (b).

*(a) Dynamics of the transformation of droplet shape*

Analysis of coalescence processes usually starts with temporal dependence of the width of bridge between droplets that allows to determine the coalescence regime. Figure 3 shows the width of the bridges $W(t)$ versus time for three pairs of droplets with different radii. The time of coalescence increases with increasing droplet size. In the inset in Fig. 3 a log-log plot is shown. On the middle stage of coalescence, power-law dependence is visible, $t^n$ where $n$ equals ¼.

Yokota and Okumura [29,33] studied merging

of a droplet with a bath of the same viscous liquid. The droplet and bath were contained in a Hele-Shaw cell filled by a low-viscous fluid. Poiseuille flow was considered. From the balance of capillarity and viscous dissipation Yokota and Okumura [29] obtained the expression for temporal

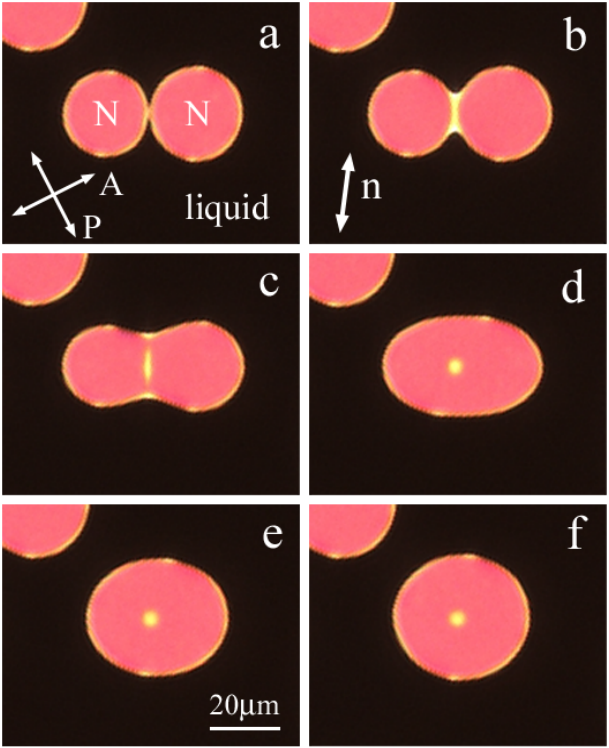

FIG. 2 (color online). Dynamics of transformation of the shape and interior structure of nematic (N) droplets in surrounding isotropic liquid. Line (yellow stripe (b)), elongated (c) and circular domains (d-f) are formed at coalescence. Thickness of the cell $h=5$ μm. Times after start of coalescence are 0.04 s (b), 0.38 s (c), 2.06 s (d), 4.26 s (e) and 8.76 s (f). The orientation of polarizer, analyzer and nematic **n**-director is shown.

dependence of bridge size. Similar consideration applied to our case of two coalescing droplets in viscous liquid gives for the temporal dependence of the bridge width

$$W(t)/2R = A(t/\tau_S)^{1/4}, \quad (1)$$

where $\tau_S = R^3\eta/h^2\gamma$, $A$ is an unknown prefactor depending in particular on the dynamics of the outer fluid, $\eta$ is the viscosity. Our data (the inset in Fig. 3) correlate with ¼ slope. Quantitative proof of this relation at present is impossible since the influence of the outer fluid viscosity and the prefactor $A$ are not known. Moreover the value of effective viscosity in E7 is known only far from the nematic-isotropic transition [34]. However it is possible to use scaling analysis. $W(t)$ was scaled by $2R$ and $t$ by $\tau_S$. Characteristic time $\tau_S$ was determined so that data followed the dependence $W(t)/2R = (t/\tau_S)^{1/4}$. Bridge dynamics in log-log scale is presented in Fig. 4(a). In the interval of bridge width $0.5 < W(t)/2R < 1$ the data follow a self-similar behavior and collapse onto a master curve [Fig. 4(a)]. The straight line shows the slope ¼.

Coalescence of two droplets on early- and middle stage of coalescence was not previously studied in a Hele-Shaw cell. A slope ¼ was found in another experiment at merging of a glycerol droplet with the bath in a Hele-Shaw cell with low-viscosity oil [29]. The characteristic time $\tau_S$ determined from scaling [circles, Fig. 4(b)] strongly depends on droplet size. The dynamics of coalescence differs from the case of coalescence of liquid lenses on a thick layer of liquid [35] and in free-standing films [36,37]. This is because the velocity gradient normal to the plane of the flow in these cases is small meanwhile in a Hele-Shaw cell it dominates [30-32].

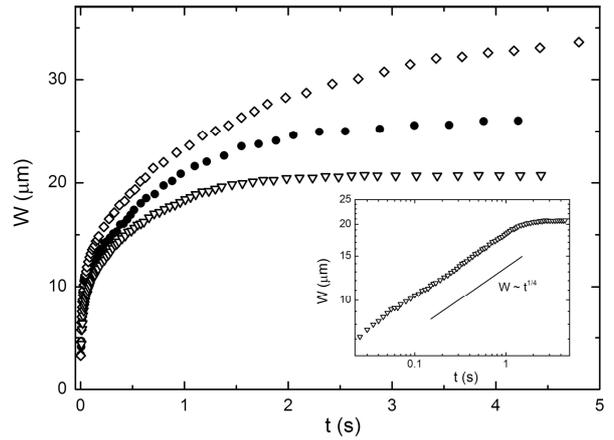

FIG. 3. The width of droplet bridge $W(t)$ versus time for nematic droplets with radius $R \approx 12.3$ μm (diamonds), $R \approx 8.9$ μm (circles), $R \approx 7.2$ μm (triangles). $h=5$ μm. The time of coalescence increases with increasing droplet size. The inset shows a log-log plot for $R \approx 7.2$ μm.

On longer timescale a secondary regime is observed. The droplet relaxation to circular shape can be described by a dimensionless parameter $D_f(t) = (L(t) - W(t))/W(t)$. Here, $L(t)$ and $W(t)$ are the droplet length and width. Brun et al. [30], Oswald and Poy [31] have shown that at the later stage of relaxation $D_f(t) = D_f(0)exp(-t/\tau_R)$. Relaxation time $\tau_R = 2^{3/2}R^3(\mu_1+\mu_2)/\alpha h^2\gamma$, where $\mu_1$ and $\mu_2$ are the viscosities of liquid and nematic, α is a dimensionless factor. $\alpha \approx 0.36$ is obtained from numerical calculations [30,31], it depends very weakly on $R/h$ and $\mu_1/(\mu_1+\mu_2)$ [30,31]. As in experiments by Oswald and Poy [31] our data for

$D_f(t)$ at the last stage follow the exponential relaxation. First determination of both $\tau_S$ and $\tau_R$ at droplet coalescence allow us to find correlation between them. Crosses in Fig. 4(b) show the relaxation time $\tau_R$. The dependence of $\tau_S$ and $\tau_R$ on $R$ and their values are close. The solid line [Fig. 4(b)] is the $\sim R^3$ dependency. So, analytical expressions both for $\tau_S$ and $\tau_R$ well describe their dependence on $R$.

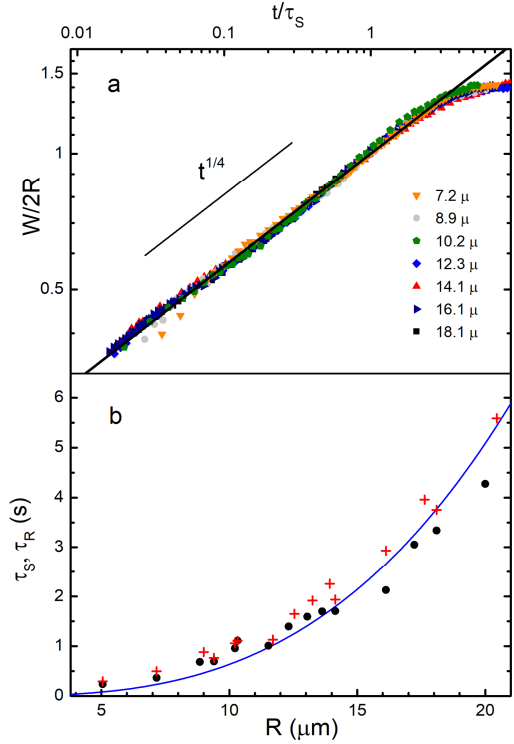

FIG. 4 (color online). (a) Log-log dependence of $W(t)/2R$ on $t/\tau_S$ for droplets from $R=7.2$ μm to $R=18.1$ μm. Data were fitted by the dependence $W(t)/R=(t/\tau_S)^{1/4}$. (b) Scaling times $\tau_S$ in the middle stage of coalescence (circles) and relaxation times $\tau_R$ for the droplet relaxation to circular shape (crosses) versus the initial droplet radius. The solid line is the power law $\sim R^3$. $h=5$ μm.

*(b) Transformation of structure inside droplets and instability of the bridge*

Geometry of the bridge strongly depends on the wetting on the cell surfaces. Figure 5 shows an image of the intermediate region between nematic (N) and liquid (L) phases. The photograph was taken in monochromatic light between crossed polarizers. The isotropic liquid looks dark, while nematic with oblique director orientation with respect to polarizers transmits light due to birefringence. Thin bright and dark stripes in the photo are related with interference in the region of the meniscus. Birefringence of E7 close to the nematic-isotropic transition is about 0.1 [38]. Shape of the meniscus determined from the position of interference stripes is shown by dots in Fig. 5(b). The meniscus shape is close to circular (the curve in Fig. 5(b) is a semicircle tangent to the cell surfaces). This observation confirms that the isotropic liquid completely wets the cell surface.

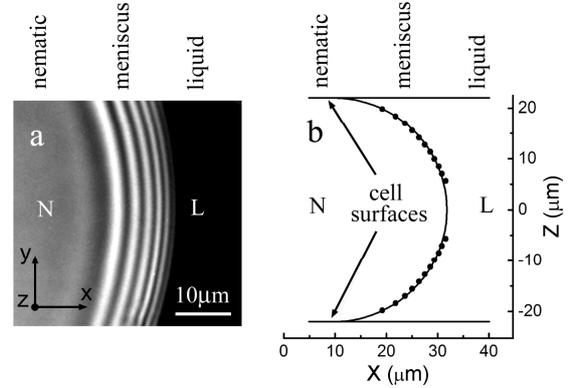

FIG 5. The intermediate region (meniscus) between nematic (N) and isotropic liquid (L) in a Hele-Shaw cell. (b) Cross-cut in the middle of the cell. Meniscus shape (dots in (b)) determined from the position of interference fringes. The curve in (b) is a semicircle. The photograph was taken with a green filter ($\lambda=545$ nm). Thickness of the cell $h=44$ μm.

Now we focus on our main result on the transformation of structure inside the droplet. Transformation of outer and droplet bridges can be divided in two stages: transformation of bridge sizes and the following instability of the outer bridge with fragmentation of the outer bridge into small droplets. Figures 6-8 show the bridge region at coalescence of large liquid (L) droplets in outer nematic (N) phase. The first stage of the transformation of bridge sizes is shown in Fig. 6, bridge fragmentation induced by the instability is illustrated in Figs. 7,8.

Figure 6 shows photos of the bridge region (front view) and schematic representation of the bridge region (side view). Observations in monochromatic light allow to characterize the shape of the bridge before coalescence [Fig. 6(a1)] and in the process of coalescence [Fig. 6(b1,c1)]. Dots in Fig. 6(a1,b1) show the shape of the bridge determined from positions of interference stripes. Before coalescence [Fig. 6(a,a1)] only the outer bridge of nematic is present. A gap for liquid exists between two liquid droplets even near the cell surfaces [Fig. 6(a1)]. As the droplets approach and touch each other the gap vanishes. When coalescence starts, due to curvature of meniscus, droplet bridge from isotropic liquid appears at the boundary of the cell. The outer bridge from nematic is localized in the middle of the cell [Fig. 6(b,b1)]. Surface tension makes the cross-section of the bridge in its middle part circular. With time both the width and the thickness of the outer bridge from nematic material essentially decrease [Fig. 6(b,c,b1,c1)]. The number of interference stripes on the left and right sides from

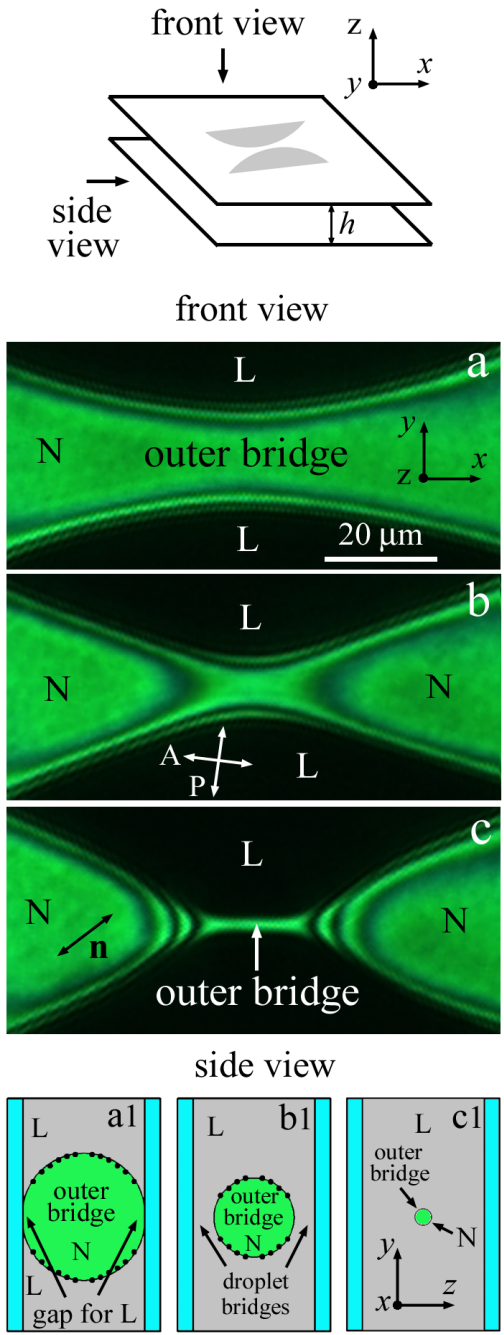

FIG. 6 (color online). Front view (a-c). Bridge region before coalescence of two isotropic droplets in nematic environment (a). Side view of the bridge region (a1-c1), cross-cut in the middle of the cell. Dots show the shape of the bridge determined from the positions of interference fringes. After start of coalescence the transverse size of the outer bridge decreases (a1-c1), the droplet bridge appears and increases (b1,c1). $R$ is about 100 μm. The photographs were taken with a green filter. $h=20$ μm.

the bridge [Fig. 6(c)] shows that the thickness of the outer bridge is about its width. Orientation of the outer and droplet bridges is perpendicular (outer bridge is along $x$-axis, droplet bridge is along $y$-axis). The existence of isotropic bridge near the cell surface is energetically favorable since liquid wets the surface (Fig. 5). Figure 7

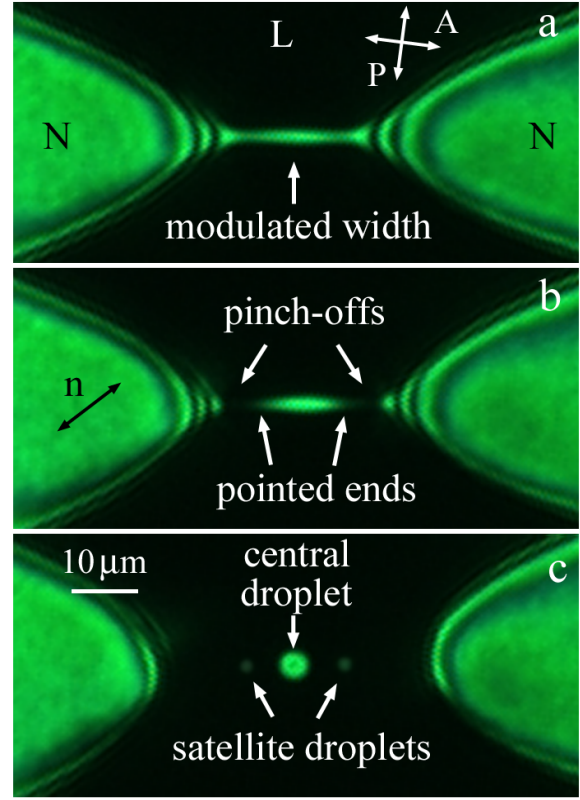

FIG. 7 (color online). Bridge region when the width of the bridge is modulated (a), pinch-off (b), central and satellite droplets (c). The photographs were taken with crossed polarizers. $h=20$ μm. Times after frame (c) in Figure 6 are 0.03 s (a), 0.07 s (b), and 0.18 s (c).

shows the coalescence process after frame (c) in Figure 6. Figure 8 in detail illustrates the second stage of transformation of the outer bridge. The photographs were taken with a green filter (Fig. 7) and in white light (Fig. 8). The second stage starts when a nearly cylindrical bridge is developed [Figs. 6(c),8(a)]. When the transverse size of the outer bridge $2r_0$ becomes significantly small (about 2.4 μm), a nearly sinusoidal modulation appears in the outer bridge [Figs. 7(a), 8(b)]. With time the modulation increases [Fig. 8(b,c)] so that the width of the outer bridge decreases near its ends and increases in the center. The length of the outer bridge increases [Fig. 8(a-c)]. Two distortions grow near the ends and symmetric pinch-offs occur in these places [Fig. 7(b),8(d)]. After pinch-off event an elongated droplet with pointed ends is formed in the center

[Figs. 7(b),8(d)]. This elongated droplet does not relax to the circular form. The droplet also becomes unstable and secondary pinch-off takes place almost at the same time at both sides of the elongated droplet. The elongated droplet splits into three domains: central droplet and two satellites [Figs. 7(c),8(e)]. The satellites as a rule are much smaller than the main central droplet. We observed

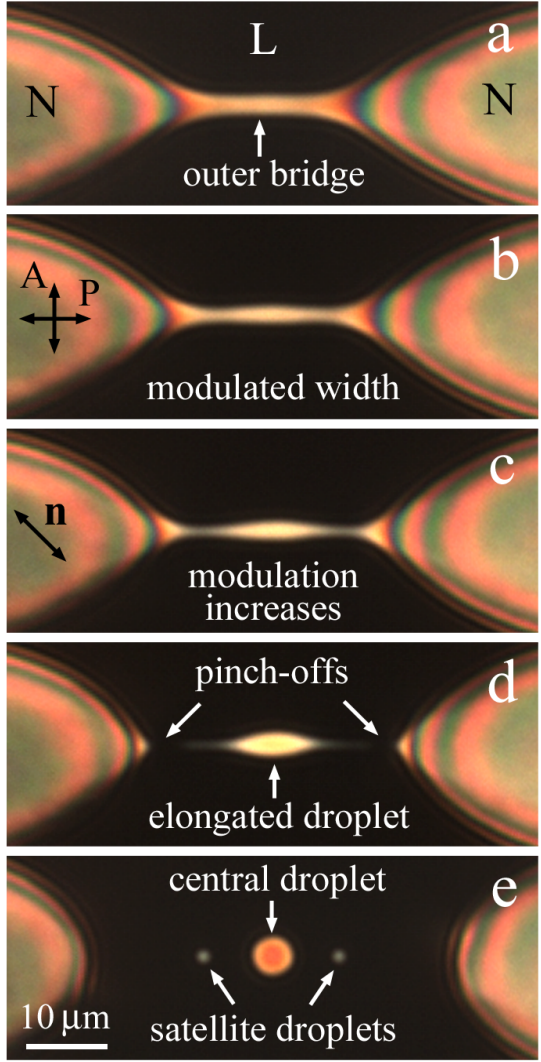

FIG. 8 (color online). Instability and breakup of the outer bridge. Modulation of the bridge width appears and increases (b,c), followed by pinch-offs with formation of an elongated droplet (d). The elongated droplet breaks and two satellite droplets form (e). The photographs were taken in white light. Time after frame (a) is 0.03 s (b), 0.06 s (c), 0.11 s (d), 0.23 s (e). $h$=20 μm.

formation of satellite droplets in cells in a wide range of thicknesses from 5 μm to 50 μm. Satellite droplets were first observed by Plateau at breakup of fluid viscous cylinders [39]. Later they were found in numerous works on breakup [16,18]. In particular, Burton and Taborek observed formation of satellite droplets at breakup of stretched liquid lenses on water surface [17]. However at coalescence of liquid lenses Burton and Taborek did not report the formation of elongated droplets and their fragmentation with formation of satellite droplets. In previous investigations of 3D coalescence and coalescence in planar cells [1-5,26,31-33] pinch-offs and satellite droplets also were not observed.

The initial small distortion of the outer bridge can result from fluctuations or small external perturbations. In our system the droplets and the environment are composed by the same material in different phases. Nematic-isotropic interfacial energy is much smaller (about $10^{-5}$ N/m) than corresponding surface energy for nematic or isotropic liquid. So, fluctuations and distortions forming thinner regions of the bridge can be much larger and become important. Our observation of the modulation, formation of the elongated droplet, central and satellite droplets are remarkably similar to the cascade of pinch-offs during breakup of 3D filaments or stretched droplets [15-20]. However let us stress once again that in our case the instability and formation of satellite droplets takes place at coalescence.

The evolution of the bridge and breakup can be compared to the theory of Rayleigh-Plateau instability of a fluid cylinder. If the wavelength $\lambda_0$ of spatial modulation of cylinder diameter is greater than $\lambda_C=2\pi r_0$, the amplitude of modulation has to grow since this decreases the surface area [16]. The growth rate of the modulation depends on its wavelength. The mode whose amplitude grows fastest leads to breakup [16]. Classical Rayleigh mode (the fastest growing mode leading to breakup) has wavelength $\lambda_R=9.01 r_0$ [16]. In our case the wavelength $\lambda_0$ is somewhat larger, about $12 r_0$. The difference can be due to several reasons. One is the finite value of viscosity which shifts the fastest growing mode to longer wavelengths [16]. Second, the above mentioned analysis is related to an infinite liquid cylinder. In our case, we deal with the breakup of a bridge of a finite size, and the length of the bridge increases during coalescence. Note that according to linear stability theory the breakup should occur at the center of the bridge [15,16]. However calculations including nonlinear terms show that breakup must occur at the ends of the bridge [15] in agreement with our experiments. This shows that the observed breakup process is essentially nonlinear.

## IV. CONCLUSION

In summary, for the first time we observed droplet coalescence and breakup simultaneously. Previously these phenomena were observed and studied in different experiments and in different sections of fluid dynamics: droplet coalescence and Rayleigh–Plateau instability inducing breakup of jets or stretched droplets. The main result of our work is finding at coalescence the instability of bridges, their pinch-off and fragmentation with formation of small satellite droplets. Such behavior at droplet coalescence earlier was not observed and theoretically was not predicted. Our work connects two important areas of fluid dynamics: coalescence and breakup. A peculiarity of the observed phenomenon is that the instability occurs not with pre-existing filaments and stretched droplets but with a thin bridge formed at the coalescence. Small nematic satellite droplets are formed at coalescence of droplets of isotropic liquid. The performed investigations broaden and deepen our knowledge and understanding of coalescence, breakup phenomena and satellite droplet formation. We also for the first time established a scaling law and correlation between characteristic times on different stages of droplet coalescence in a Hele-Shaw cell.

This work was supported by the Russian Science Foundation under grant 18-12-00108.